\documentclass[conference]{IEEEtran}
\usepackage{amssymb}
\usepackage{amsmath}
\usepackage{amsfonts}
\usepackage{graphicx}
\usepackage{color}
\usepackage{algo}
\def\booknames#1/#2/#3/#4{#1:#2:#3:#4}
\newtheorem{defi}{Definition}
\newtheorem{prop}{Proposition}

\IEEEpeerreviewmaketitle

\begin{document}
\title{SNAP: SNowbAll multi-tree Pushing for Peer-to-Peer Media Streaming}
\author{
\IEEEauthorblockN{Jun Luo}
\IEEEauthorblockA{
    School of Computer Engineering, Nanyang Technological University (NTU), Singapore \\\texttt{junluo@ntu.edu.sg}}
}
\maketitle

\begin{abstract}
  Given the respective advantages of the two complimentary techniques for peer-to-peer media streaming (namely \textit{tree-based push} and \textit{mesh-based pull}), there is a strong trend of combining them into a hybrid streaming system. Backed by recently proposed mechanisms to identify \textit{stable peers}, such a hybrid system usually consists of backbone trees formed by the stable peers and other overlay structures in the second tier to accommodate the remaining peers. In this paper, we embrace the hybrid push-pull structure for peer-to-peer media streaming. Our protocol is dominated by a multi-tree push mechanism to minimize the delay in the backbone and is complemented by other overlay structures to cope with peer dynamics. What mainly distinguishes our multi-tree pushing from the conventional ones is an unbalanced tree design guided by the so called \textit{snow-ball streaming}, which has a provable minimum delay and can be smoothly ``melded" with virtually any other existing overlay structures lying in the second tier. We design algorithms to construct and maintain our \textit{SNowbAll multi-tree Pushing} (SNAP) overlay, and we also illustrate how to smoothly weld the SNAP backbone with the second tier. Finally, we perform simulations in \textit{ns}-2; the results indicate that our approach outperforms a recently proposed hybrid streaming system.
\end{abstract}


\section{Introduction} \label{sec:intro}
  Although the tremendous growth of peer-to-peer (P2P) streaming applications on the Internet has attracted a great number of users and also made the P2P streaming a heavily investigated field, there are still major obstacles in its adoption as a mainstream and commercialized broadcast services \cite{LiuRLZ-IEEEProc07}. One of the major technique obstacles, as pointed out by \cite{LiuRLZ-IEEEProc07,HeiLR-IEEECOMMAG08,FengLL-INFOCOM09}, is the lack of guarantee on the \textbf{delay} performance.

  There are two mainstream approaches to deploy a P2P streaming system, namely \textit{tree-based push} (e.g., \cite{BanerjeeBK-SIGCOMM02,CastroDKNRS-SOSP03,VenkataramanYF-ICNP06}) and \textit{mesh-based pull} (e.g., \cite{ZhangLLY-INFOCOM05,HeiLR-IEEECOMMAG08,LiXQKLLZ-INFOCOM08}). Both approaches have their pros and cons. While the tree-based push achieves better performance in terms of delay but suffers higher maintenance complexity and data outage upon peer dynamics \cite{MaghareiRG-INFOCOM07}, the mesh-based pull provides a higher robustness against peer dynamics but has to strike a compromise between efficiency and latency \cite{FengLL-INFOCOM09}. Recently, several proposals started to promote a hybrid approach \cite{WangXL-ICDCS07,ZhangZSY-IEEEJSAC07,WangLX-INFOCOM08}; they all aim at reaching a balance between the two approaches above.

  Unfortunately, directly melding a conventional tree overlay with a mesh overlay may not yield the optimal performance. The difficulty stems from a fundamental difference between the push and pull approaches in terms of handling the media streams. More specifically, while a push approach usually treats the streams at the \textit{packet} level, a pull approach organizes the packets of a stream into larger units (sometimes referred to as \textit{chunks}) as pulling individual packets would result in a much higher overhead. A direct consequence of this difference is that a hybrid approach has to take the chunk as its data unit in order to be compatible with its pull components. An immediate question about this chunk-oriented hybrid system is the following: is it proper to use an optimal packet-oriented push tree (e.g., \cite{LiuSJRC-SIGMETRICS08}) as the push component of the hybrid approach? The answer is \textbf{negative}, as we will illustrate in Sec.~\ref{sec:best}. This has motivated us to identify the optimal push tree component for a hybrid P2P streaming system.

  It is well known that the minimum delay to disseminate a \textbf{single} chunk to $N$ peers with homogeneous uploading capacity is $1+\lceil \log_2 N\rceil$ and it can be achieved by a \textit{snow-ball streaming} algorithm \cite{Liu-MM07}. However, only the existence of such a mechanism for \textbf{streaming} has been shown in \cite{Liu-MM07}; the lack of a systematic way to construct a distributed scheduling for this algorithm has significantly confined its applicability. Existing proposals that aim at minimizing delay have to rely on certain approximation mechanisms, either deterministic pull \cite{FengLL-INFOCOM09} or randomized push \cite{BonaldMMPT-SIGMETRICS08}. One of the main targets of our paper is to bring forth a straightforward mapping from the snow-ball algorithm to a multi-tree overlay. This will greatly facilitate the distributed minimum-delay chunk scheduling in a push-pull hybrid streaming system.

  In this paper, we investigate the chunk-oriented hybrid approach for P2P media streaming. We advocate a hierarchical structure where stable peers \cite{WangLX-INFOCOM08,LiuWLZ-INFOCOM09} are organized into a multi-tree backbone while the remaining peers may either pull chunks from the backbone or further organize themselves into sub-trees attached to the backbone. The main contributions of our paper is the following:
  \begin{itemize}
  \item We propose algorithms to construct and maintain a multi-tree overlay. The scheduling policy guided by these trees guarantees a minimum delay for continuous chunk streaming, as promised by the snow-ball streaming algorithm (which only shows the existence of such a policy).
  \item We design a P2P streaming backbone, called \textit{SNowbAll multi-tree Pushing} (SNAP), based on the proposed multi-tree overlay. We also demonstrate how to build a hybrid push-pull streaming system by flexibly combining SNAP with other overlay structures.
  \item We implement a streaming system in \textit{ns}-2 by combining SNAP with PPLive \cite{HeiLLLR-IPTV06}, our simulation results show that SNAP+PPLive outperforms its up-to-date competitor.
  \end{itemize}

   In the next section, we provide detailed explanations on why the snowball streaming algorithm, rather than a traditional tree-based push, has to be used for chunk streaming. We then describe in Section~\ref{sec:snap} the algorithms that construct and maintain SNAP. In Section~\ref{sec:hybpp}, we extend SNAP into a full-fledged P2P streaming system by  welding it with various overlay structures that serve as the second tier of the system. We report the experiment results in Section~\ref{sec:perfevl} and survey the related work in Section~\ref{sec:rw}. Finally, we conclude our paper and discuss our future work in Section~\ref{sec:con}.

\section{What Is The Best Tree Overlay?} \label{sec:best}
  In this section, we investigate the feature of the chunk-oriented streaming. By quantitatively comparing two tree overlays that are claimed to be optimal under different circumstances, we show that common wisdom valid in packet-oriented streaming does not yield the optimal performance under the chunk-oriented system.

  The major difference between a packet-oriented and a chunk-oriented system lies in the concept of delay. For data transmission across the Internet, there are basically two main components of the delay: \textit{transmission delay} $t$ and \textit{queueing}/\textit{propagation delay} $d$. In a packet-oriented system, $t \ll d$ in general, so the delay is dominated by $d$. The same situation does not apply to a chunk-oriented system, as a chunk may aggregate several hundreds of packets and hence $t \approx d$.

\begin{figure}[!h]
   \begin{center}
        \includegraphics[width=\columnwidth]{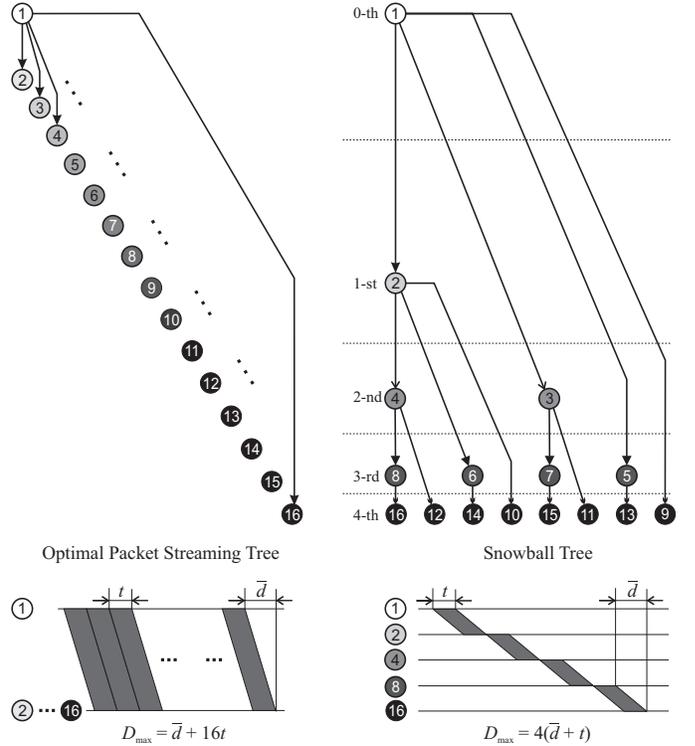}
    \caption{The comparison between the \textit{optimal packet streaming tree} (OPST) \cite{LiuSJRC-SIGMETRICS08} and the \textit{snowball tree} (SBT) \cite{Liu-MM07} in terms of disseminating one packet or chunk in a system of 16 peers. The background color of a peer indicates its delay: the darker the color the longer the delay is. For the SBT, we also illustrate the \textit{peer level set} defined later in the paper.}\label{fig:treecomp}
   \end{center}
  \end{figure}

  Now let us see how this difference affects on the optimal tree overlay. To this end, we compare two tree overlays: the \textit{optimal packet streaming tree} (OPST) \cite{LiuSJRC-SIGMETRICS08} and the \textit{snowball tree} (SBT) \cite{Liu-MM07}\footnote{The snowball streaming algorithm, as appeared in \cite{Liu-MM07}, does not involve a tree structure, but it can be easily mapped to a tree in terms of disseminating one chunk. What remains challenging is how to map the algorithm to a multi-tree structure for chunk streaming. We will solve the problem in Sec.~\ref{sec:snap}.}. As shown in \cite{LiuSJRC-SIGMETRICS08}, OPST is indeed the optimal tree for packet streaming, so we omit the comparison with other tree overlays (e.g. \cite{CastroDKNRS-SOSP03}). To simply the interpretation, we only compare the two structures in terms of disseminating one packet or chunk. The extension to streaming multiple packets or chunks is described in \cite{LiuSJRC-SIGMETRICS08} (for OPST) and will be introduced in Sec.~\ref{sec:snap} (for SBT). Let \framebox{$\mathcal{N}$} be the set of all peers and $|\mathcal{N}| = N$, $\bar{d}$ be the average queueing/propagation delay, and $t$ be the transmission time of a packet or a chunk. Fig.~\ref{fig:treecomp} illustrates the two tree overlays in a system of $N=16$.
  Note that, for SBT, we organize peers into levels. We denote the peer that directly receives a chunk from the server as the 0-th level, and peers that are at the $k$-th level are those that receive the chunk in the $k$-th round of the geometric progression\footnote{The snowball streaming algorithm \cite{Liu-MM07} proceeds in a round-by-round manner: a peer that receives a chunk in a round will keep pushing the chunk to others during the later rounds until everyone gets it. Therefore, if we use $x(k)$ to represent the number of peers that have the chunk at the end of the $k$-th round, $x(k) = 2^k$ is a geometric progression with common ratio 2.} of the snowball streaming algorithm.

  For OPST, after one peer (peer 1 in this case) obtains a packet or chunk from the server, it sequentially sends the packet or chunk to the rest of the peers. This leads to the following maximum delay $D_{\mathrm{max}}$ and average delay $\bar{D}$.
  \begin{eqnarray}
  D_{\mathrm{max}}^{\mathrm{OPST}} &=& \bar{d} + Nt \\
  \bar{D}^{\mathrm{OPST}}          &=& \bar{d} + 0.5(N+1)t
  \end{eqnarray}
  Note that, as shown in \cite{Liu-MM07}, parallel sending would lead to the same maximum delay but higher average delay. Also, we deliberately omit the delay of peer 1 downloading from the server, as this is a constant that contributes to the delay of every peer. The reason why $\bar{d}$ only contributes to the delay once is exactly due to the well known \textit{pipelining} effect in the Internet, as also illustrated in Fig.~\ref{fig:treecomp}. Although pipelining traditionally refers to the streaming along the same transmission path, the same effect applies here even though the destinations are different, because the whole system appears like a pipe to the sender (peer 1).

  For SBT, every peer keeps pushing the received packets or chunks to other peers, which results in very different transmission paths as shown in Fig.~\ref{fig:treecomp}. The pipelining effect appears only in certain paths, leading to a very different expressions of the delays.
  \begin{eqnarray}
  D_{\mathrm{max}}^{\mathrm{SBT}} &=& \lceil\log_2 N\rceil \left(\bar{d} + t\right) \label{exp:maxmindelay} \\
  \bar{D}^{\mathrm{SBT}}          &\leq& \frac{\lceil\log_2 N\rceil}{2}\bar{d} + \left(\lceil\log_2 N\rceil - 1\right) t + o(N) \label{exp:avgmindelay}
  \end{eqnarray}
  The maximum delay comes from the path for which there is no pipelining, as illustrated in Fig.~\ref{fig:treecomp} (following the path $1 \rightarrowtail 2 \rightarrowtail 4 \rightarrowtail 8 \rightarrowtail 16$. The derivation of the average delay is a bit tricky and we postpone it to Appendix~\ref{sec:avg}. Note that our derivation differs from the one presented in \cite{Liu-MM07} in that we explicitly separate the two delay components.

  Given all the delay expressions, it is obvious that, when $t \approx \bar{d}$ (or $t > \bar{d}$) and $N$ is relatively large, SBT yields a much smaller delay (both maximum and average) than OPST. In the extreme case where $d\rightarrow0$, we know that SBT is indeed the optimal tree according to \cite{Liu-MM07}.

\section{Snowball Pushing by a Multi-SBT Structure} \label{sec:snap}
  Although SBT is shown to be superior to OPST for single chunk dissemination, the extension of SBT for multi-chunk dissemination (hence media streaming) is far less trivial then the extension of OPST. As shown in \cite{LiuSJRC-SIGMETRICS08}, extending OPST for multi-chunk dissemination can be achieved simply by a multi-OPST structure: it consists of $N$ trees, where the root of the $i$-th tree is peer $i$, and the server simply pushes the packets to these trees in a round-robin fashion. Unfortunately, Such an extension obviously does not apply to SBT, as both the internal and leaf peers in SBT need to be careful rescheduled. Recently, Trellis graphs are used to represent the snowball streaming algorithm \cite{FengLL-INFOCOM09}, but it fails to suggest a distributed chunk scheduling policy. Therefore, we need to find a multi-tree extension for SBT that provides the same delay guarantee for chunk streaming. To this end, we first propose a centralized construction algorithm, then we show how to make it operate in a distributed manner.

\subsection{Multi-Tree Representation of The Snowball Algorithm}
  In the original proposal of the snowball streaming algorithm \cite{Liu-MM07}, the existence of an algorithm for chunk streaming is proven by an induction, which cannot be used to derive a multi-tree structure. We hereby propose a centralized way to construct a multi-tree representing the optimal scheduling policy. More precisely, we have the following design goal:
  \begin{quote} \textit{A set of SBTs such that, if the server takes turn to push chunks to them in a round-robin fashion, the minimum delay is achieved for every chunk.}
  \end{quote}
  For practical purposes, we require the multi-tree structure to consists of a finite number of SBTs; in other words, the chunk dissemination follows a repetitive pattern of trees with a period of \framebox{$P$}. Note that, as each chunk has a corresponding SBT and we know that SBT is delay optimal for single chunk dissemination, the challenge we are facing is to resolve the parallelism among all these $P$ SBTs.

  We refer to the $i$-the SBT in this $P$-tree pattern as \framebox{$T_i$}. We denote by \framebox{$L^p_{k,i}$} the \textit{peer level set} containing the set of peers in the $k$-th level of $T_i$, and by \framebox{$L^e_{k,i}$} the \textit{edge level set} containing the edges whose ends are in $L^p_{k,i}$ .  We illustrate the concept of peer level set in Fig.~\ref{fig:treecomp}, where we only show an arbitrary $T_i$ with $\{L^p_{0,i}, L^p_{2,i}, L^p_{2,i}, L^p_{3,i}, L^p_{4,i}\}$ and with $L^p_{0,i}$ containing the peer that directly receives a chunk from the server. Finally, let $K = \lceil\log_2 N\rceil$ be the maximum depth of every SBT. As the minimum delay can be guaranteed for every SBT only if the maximum parallelism is achieved in terms of the uploading scheduling, we first make the following observations on the concurrent scheduling of the uploading edges of the SBTs.
  \vspace{1ex}
  \begin{prop} \label{prop:feasible}
    The following conditions constrain the possible scheduling of the uploading along the tree edges:
    \begin{enumerate}
    \item Edges in the the same tree can be scheduled at the same time iff they belong to the same level set: If $l_1, l_2 \in T_i$, then $l_1,l_2 \in L^e_{k,i}$ for $1 \leq k \leq K$ $\boldsymbol{\Longleftrightarrow} l_1 \parallel l_2$.
    \item Edges in different trees can be scheduled at the same time iff their do not share the same origin and they belong to two level sets whose level indices differ less than the difference in tree indices: If $l_1, \in T_i$, $l_2 \in T_j$, and $i - j = a~ (\!\!\!\!\mod P)$, then $\left(l_1, \in L^e_{k_1,i}\right) \wedge \left(l_2 \in L^e_{k_2,j}\right) \wedge \left(k_2 - k_1 \leq a\right) \wedge \left(o(l_1) \neq o(l_2)\right) \boldsymbol{\Longleftrightarrow} l_1 \parallel l_2$, where $o(l)$ represents the origin of edge $l$.
    \end{enumerate}
  \end{prop}
  We prove this proposition in Appendix~\ref{sec:tsch}. In the following, we give a sufficient condition for a multi-tree structure to yield the minimum delay for chunk streaming.
  \vspace{1ex}
  \begin{prop} \label{prop:sufficient}
    The following condition guarantees a multi-tree structure to yield the minimum delay: If the edges in $L^e_{1,i}$ is scheduled, then all edges in $L^e_{k,i-k+1}:k\in(2,K)$, as well as the $N-2^{K-1}$ edges in $L^e_{K,i-K+1}$, should be able to be scheduled together.
  \end{prop}
  \vspace{1ex}
  We prove this proposition in Appendix~\ref{sec:mdmtree}.

  To facilitate further discussion, we use the following definition to characterize the \textbf{origins} of the edges involved in the schedule described in \textit{Proposition~\ref{prop:sufficient}}.
  \vspace{1ex}
  \begin{defi}
  A set of peers is an \textit{independent set} (ISet) if they are the origins of all the edges involved in the schedule described in \textit{Proposition~\ref{prop:sufficient}}. In particular, for some integer $K$, a $K$-ISet refers to an ISet in a system of $N=2^K$ peers.
  \end{defi}
  \vspace{1ex}

  According to \textit{Proposition~\ref{prop:feasible}} (condition 2 in particular), an ISet should not contain duplicate peers. In the next section, we will explain how we can construct a Multi-SBT structure that satisfies the requirement set in \textit{Proposition~\ref{prop:sufficient}}. We will also use Fig.~\ref{fig:sptrees} to give a more tangible illustration of the concepts of ISets, as well as the sufficient scheduling required by \textit{Proposition~\ref{prop:sufficient}}.

\subsection{Constructing the Multi-SBT Structure} \label{sec:csbt}
  Given the characteristics of feasible schedules stated in \textit{Proposition~\ref{prop:feasible}} and the sufficient condition to achieve the minimum delay described in \textit{Proposition~\ref{prop:sufficient}}, our design goal is achieved if we could identify a multi-tree structure that satisfies all the conditions demanded in both propositions. We first give, in Fig.~\ref{fig:algomtree}, the algorithm that performs the construction for $N=2^K$, then we show its extension to an arbitrary $N$. Their correctness is proven in Appendix~\ref{sec:algomtree}.
  \begin{figure}[htb]
  \begin{algorithm}{Multi-SBT Construction for $N=2^K$}{
  \label{algo:algostd}}
  $\mathcal{S} \qlet \mathcal{N};~~s_k \qlet \emptyset, k=0,1,\cdots,K;~~k \qlet 0$ \\
  \qrepeat \\
    $s_k \qlet (K-k)2^{(k-1)^+}$ distinct peers in $\mathcal{S}$ \\
    \qdo assign the peers in $s_k$ to the $k$-th level of the SBTs in a periodic fashion. \\
    $\mathcal{S} \qlet \mathcal{S} \backslash s_k;~~k \qlet k + 1$
  \quntil $k = K$ \\
  Fill the $L^p_{K,i}$ with the peers that have not appeared in $T_i$.\\
  \qreturn $P$ distinct SBTs and $\{s_0, s_1, \cdots, s_K\}$.
  \end{algorithm}
  \caption{The centralized multi-tree construction algorithm for $N=2^K$} 
  \label{fig:algomtree}
  \end{figure}

\begin{figure*}[t]
   \begin{center}
        \includegraphics[width=\textwidth]{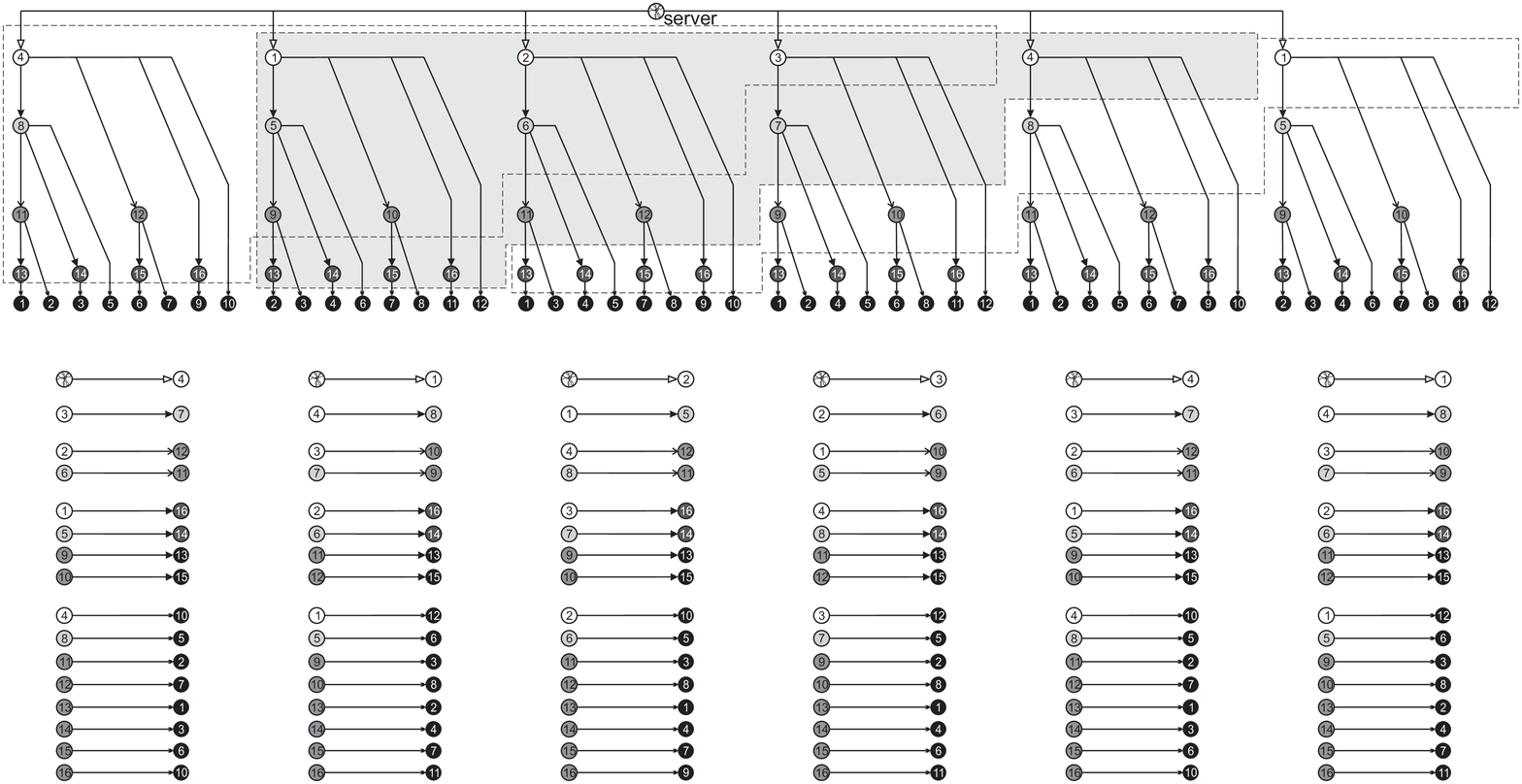}
    \caption{The multi-SBT structure, $K$-ISet, and corresponding uploading schedules for $N=16$. The correspondence between an SBT and the scheduled uploadings below it is such that these concurrent uploadings take place exactly when the server pushes a chunk to the root of this tree.} 
    \label{fig:sptrees}
   \end{center}
  \end{figure*}

  Basically, the algorithm starts with empty ``skeleton" of trees whose vertices need to be indexed. The data structure used to represent the trees is $\{s_k\}_{k=0,1,\cdots,K}$, with $s_k = \bigcup_i L^p_{k,i}$ (i.e., $s_k$ includes the peer level sets of all the trees). According to the 3rd and 4th steps of the algorithm, the peers in the $k$-th level of a multi-SBT structure repeat in a  period of $P_k = K-k$, given the number $|s_k|$ of peers to be put in the $k$-th level of the multi-SBT structure and the fact that there are $2^{(k-1)^+}$ peers in $L^p_{k,i}$ of a tree $T_i$. Consequently, the \textbf{maximum} number of SBTs required is the \textit{least common multiple} (LCM) of a set of periods $\{K, K-1, \cdots, 1\}$. Fortunately, the period $P$ of the tree pattern can actually be much smaller than that number. As explained in Appendix~\ref{sec:algomtree}, the total number of peers required to complete the algorithm is $2^K-1=N-1$. Therefore, we can make use of this extra peer to increase the period of the first level to $K$, which has the potential to greatly reduce $P$. We illustrate the outcome of this algorithm for $N=16$ in Fig.~\ref{fig:sptrees}: the ISets (in particular 4-ISets) are shown as the sets of peers that encompassed in the staircase-shaped frames. Although the LCM of the set of periods $\{4,3,2,1\}$ is 12, the actual period is just 4, as we use that extra peer to increase $P_1$ from 3 to 4.

  For an arbitrary $N$ where $2^{K-1} < N < 2^K$, the size of an ISet is between $2^{K-1}$ and $2^K$. Therefore, we simply put $N-2^{K-1}$ peers in arbitrary positions in the relative complement of (K-1)\mbox{-ISet} in K\mbox{-ISet}, which is effectively equivalent to changing the periods at certain levels, as shown by the steps from 4 to 7 of the algorithm in Fig.~\ref{fig:algomtreeplus} (which is an extension of the basic algorithm in Fig.~\ref{fig:algomtree}). These extra peers are then used to further upload chunks to the incomplete $L^p_{K,i}$ (which contains only $N-2^{K-1}$ peers) in every $T_i$.
  \begin{figure}[htb]
  \begin{algorithm}{Multi-SBT Extension for Arbitrary $N$}{
  \label{algo:algoext}}
  Choose arbitrary $\widetilde{\mathcal{N}} \subseteq \mathcal{N}$ s.t. $|\widetilde{\mathcal{N}}| = 2^{\lfloor\log_2 N \rfloor}$ \\
  Run Multi-SBT Construction for $\widetilde{\mathcal{N}}$ \\
  $\mathcal{S} \qlet \mathcal{N} \backslash \widetilde{\mathcal{N}};~~s \qlet \emptyset;~~k \qlet 0$ \\
  Choose an ascendingly ordered set $\mathcal{K} = \{k_1, k_2, \cdots, k_m\}$ s.t. $|\mathcal{S}| = \sum_{k\in\mathcal{K}} 2^{(k-1)^+}$ \\
  \qfor \textbf{all} $k\in\mathcal{K}$ \\
    $s \qlet 2^{(k-1)^+}$ distinct peers in $\mathcal{S};~~s_k \qlet s_k \cup s$ \\
    \qdo assign the peers in $s_k$ to the $k$-th level of the SBTs in a periodic fashion;~~$\mathcal{S} \qlet \mathcal{S} \backslash s$ \qrof\\
  Fill the $L^p_{K,i}$ with the peers that have not appeared in $T_i$.\\
  \qreturn $P$ distinct SBTs and $\{s_0, s_1, \cdots, s_K\}$.
  \end{algorithm}
  \caption{The centralized multi-tree extension algorithm for an arbitrary $N$} \label{fig:algomtreeplus}
  \end{figure}
  For example, given $N=20$, we have many strategies to extend from the case of $N=16$. These may include:
  \begin{enumerate}
  \item Increase $P_0$, $P_1$, and $P_2$ all by 1, i.e., $\mathcal{K} = \{0,1,2\}$.
  \item Increase $P_3$ to 2: i.e., $\mathcal{K} = \{3\}$.
  \item Replace any 4 peers in the fourth level by a repetitive pattern: $\mathcal{K} = \{4\}$.
  \end{enumerate}
  as well as any arbitrary combinations of the above strategies that increase the size of an ISet by 4 peers. The 4 extra peers added to an ISet are then used to further upload chunks to the partial $L^p_{5,i}$ (including only 4 peers) appended to every $T_i$ in the 8th step of the algorithm.

\subsection{Distributed Snowball Pushing}
  The centralized multi-SBT construction algorithms that we have proposed can only be executed by the server. To be practical for P2P media streaming, our \textit{SNowball multi-tree  Pushing} (SNAP) based on the multi-SBT should operate without each peer knowing the global information about all the SBT trees. More precisely, we need to answer the following three questions:
  \begin{itemize}
  \item [Q1:] What information should be contained in the neighbor table of a peer?
  \item [Q2:] How to efficiently reconstruct the SBTs upon peer joining or leaving?
  \item [Q3:] How to deal with bandwidth heterogeneity?
  \end{itemize}

\subsubsection{Sizing the Neighbor Table}
  A complete neighbor table of a peer should contain the children of this peer in different SBTs; it consists of subsets of peers for different trees. For example, as shown in Fig.~\ref{fig:sptrees}, the neighbor table of peer 1, $\{[5,10,16,12]\}$, has only one subset of peers, while that of peer 11 contains two subsets, $\{[13,3],[13,2]\}$. Having such a complete neighbor table at every peer, SNAP can proceed in a distributed way: each peer, upon receiving a new chunk, chooses a subset of peers in a \textbf{round-robin} fashion and pushes the chunk to the children in the subset \textbf{sequentially}. It is straightforward to see that, for a single SBT (as shown in Fig.~\ref{fig:treecomp}), the size of a neighbor table is bounded by $\lceil\log_2 N\rceil$, where the bound is obtained at the root. The periodic rotation of peers in every level of the multi-SBT structure inevitably increases this size, but, as shown by the following proposition, this increase is not drastic.
  \vspace{1ex}
  \begin{prop} \label{prop:nbsizebd}
    The size of the largest neighbor table (owned by the root of each SBT) is $\mathcal{O}\left(\lceil\log_2 N\rceil^2\right)$.
  \end{prop}
  \vspace{1ex}
  We sketch the proof of this proposition in Appendix~\ref{sec:bndntb}. In practice, peers may have a neighbor table of much smaller size. For example, peers in the $(K-2)$-th and $(K-1)$-th levels only have a neighbor table of size bounded by 3.\footnote{The same peer appears in different subsets is only counted once, as the information about this peer only occupies a single memory block, while each subset containing the peer uses a reference to refer to this memory block.} This shows that SNAP runs correctly with a neighbor table much smaller than $\mathcal{N}$. In other words, its scalability is guaranteed.

\subsubsection{Coping with Peer Dynamics} \label{sec:peerdy}
  From the server point of view, peer joining and leaving simply reshape the multi-SBT overlay. For peer join, this reshaping is conducted by an algorithm similar to what is shown in Fig.~\ref{fig:algomtreeplus}; the basic idea is to adapt to the variation in $N$ by adjusting the periods at certain levels. The reshaped multi-SBT is then conveyed to certain peers by updating their neighbor tables.

  If a peer leaves SNAP, its starved children in certain SBTs will alert the server. Upon receiving such alerts, the server will assign a parent for these peers at a lower level and also (locally) update their neighbor tables, which actually reduces the level of the corresponding peers. Let us consider the SNAP shown in Fig.~\ref{fig:sptrees}, if peer 5 leaves, peers 9, 14, 6 in the first SBT will be starved. As a response to the alerts from them, the server will replace 5 by 9 in the first tree and replace 9 by 13 in both first and third trees. Of course, all the edges (uploadings) ending at 5 are removed. It can be easily shown that such local replacements can always maintain the optimality of the multi-SBT structure. It is true that the delay will be increased during this repairing phase, which actually accounts for the fact that the delay obtained in our simulations is not optimal (see Sec.~\ref{sec:perfevl}). However, the same simulation results also show that SNAP performs much better than the most up-to-date competitor \cite{WangLX-INFOCOM08}, due to the use of the optimal pushing tree and the fact that repairing is always needed for a structured streaming overlay. Also, as we will show in Sec.~\ref{sec:hybpp}, SNAP, which severs as the backbone of our hybrid push-pull streaming system, consists of only stable peers. Therefore, we expect the repairing to be rare events. Moreover, the hybrid mechanism involved in the extended system design allows peers to pull lost chunks during the reshaping phases.

\subsubsection{Heterogeneous Cases} \label{sec:hetero}
  The biggest disadvantage of snowball streaming is its limited compatibility with the heterogeneity in peers' uploading bandwidth. As shown in \cite{Liu-MM07}, apart from two very special cases, it is generally impossible to perform scheduling for heterogeneous cases.

  To get around this limitation, we propose to unify the bandwidth by ``slicing" (through time sharing) the uploading capacity of individual peers. More precisely, for a peer $i$ to joint SNAP, it must at least offer a \textit{baseline uploading bandwidth} $B_\mathrm{base}$, i.e., $B_i \geq B_\mathrm{base}$, where $B_i$ is the uploading bandwidth of peer $i$; otherwise the peer has to stay out of the SNAP and be accommodated by a second tier overlay, as will be discussed later. The value of $B_\mathrm{base}$ is usually defined by the feature of the streamed media. For example, $B_\mathrm{base} = 300$Kbps if the streaming rate is 300Kbps. After joining SNAP, $B_\mathrm{base}$ is sliced from $B_i$, while the remaining bandwidth $B_i - B_\mathrm{base}$ (if still non-zero) can be used in several different ways depending on its quantity. Details on making use of this extra bandwidth, as well as on accommodating the non-stable or low-bandwidth peers, will be discussed when we present our extended system design based on SNAP in Sec.~\ref{sec:hybpp}.

  We note that our design entails a hybrid and hierarchical system: as not all peers are able to offer $B_\mathrm{base}$, those whose uploading bandwidth is scarce need to joint the second tier. They may pull chunks from the server, peers in the SNAP backbone, or among each other, or they may join (sub)tree streaming but receive media contents with lower rate. Those peers inevitably suffer the efficiency-latency tradeoff (for mesh pull, due to the reason explained in \cite{FengLL-INFOCOM09}) or lower quality (for tree push, due to the lower bandwidth offered by such peers), but this is the price that a system has to pay for accommodating heterogeneity.

\section{P2P Media Streaming Using SNAP} \label{sec:hybpp}
  \begin{figure*}[ht]
   \begin{center}
       \parbox{\textwidth}{\parbox{.33\textwidth}{\center\includegraphics[width=.295\textwidth]{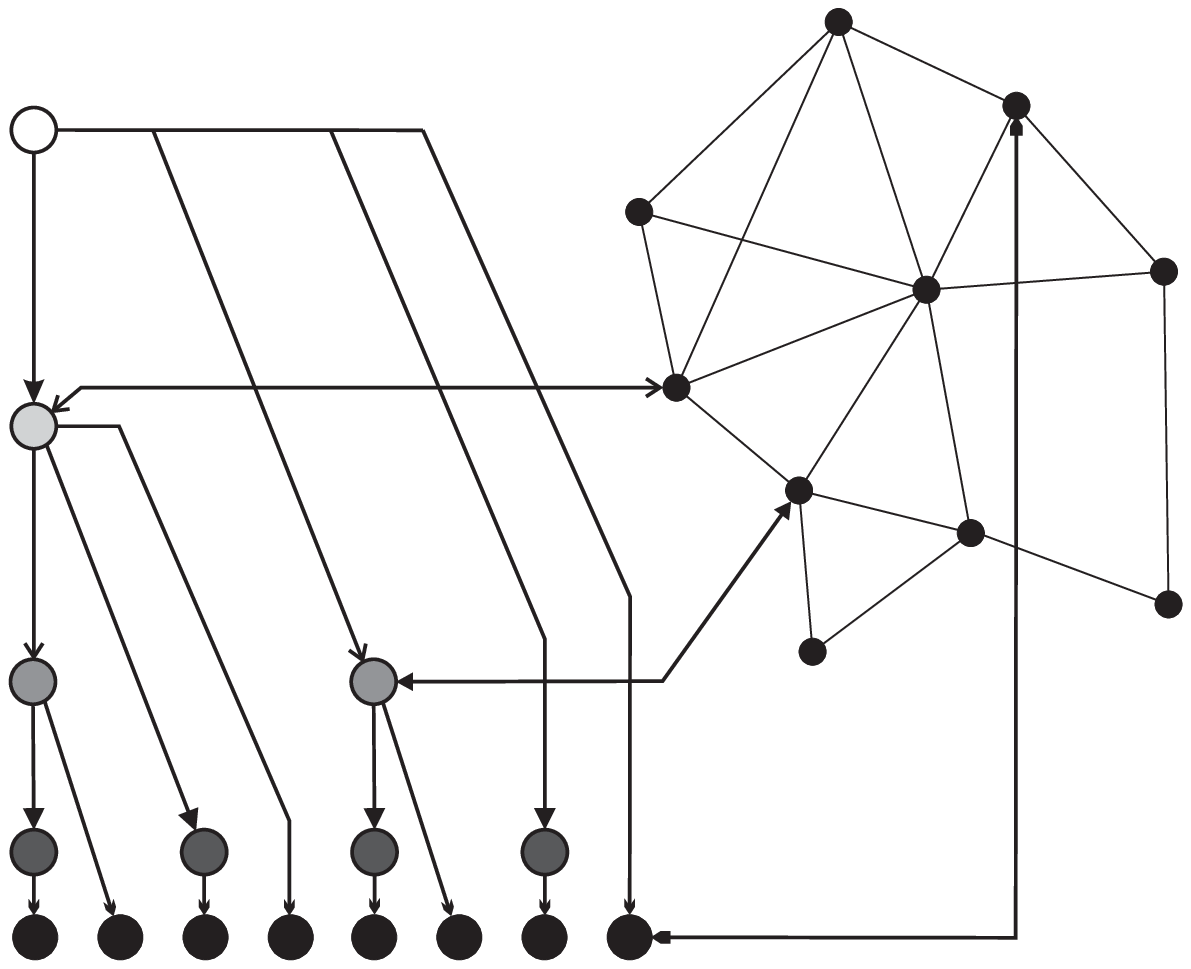}}
                           \parbox{.33\textwidth}{\center\includegraphics[width=.3\textwidth]{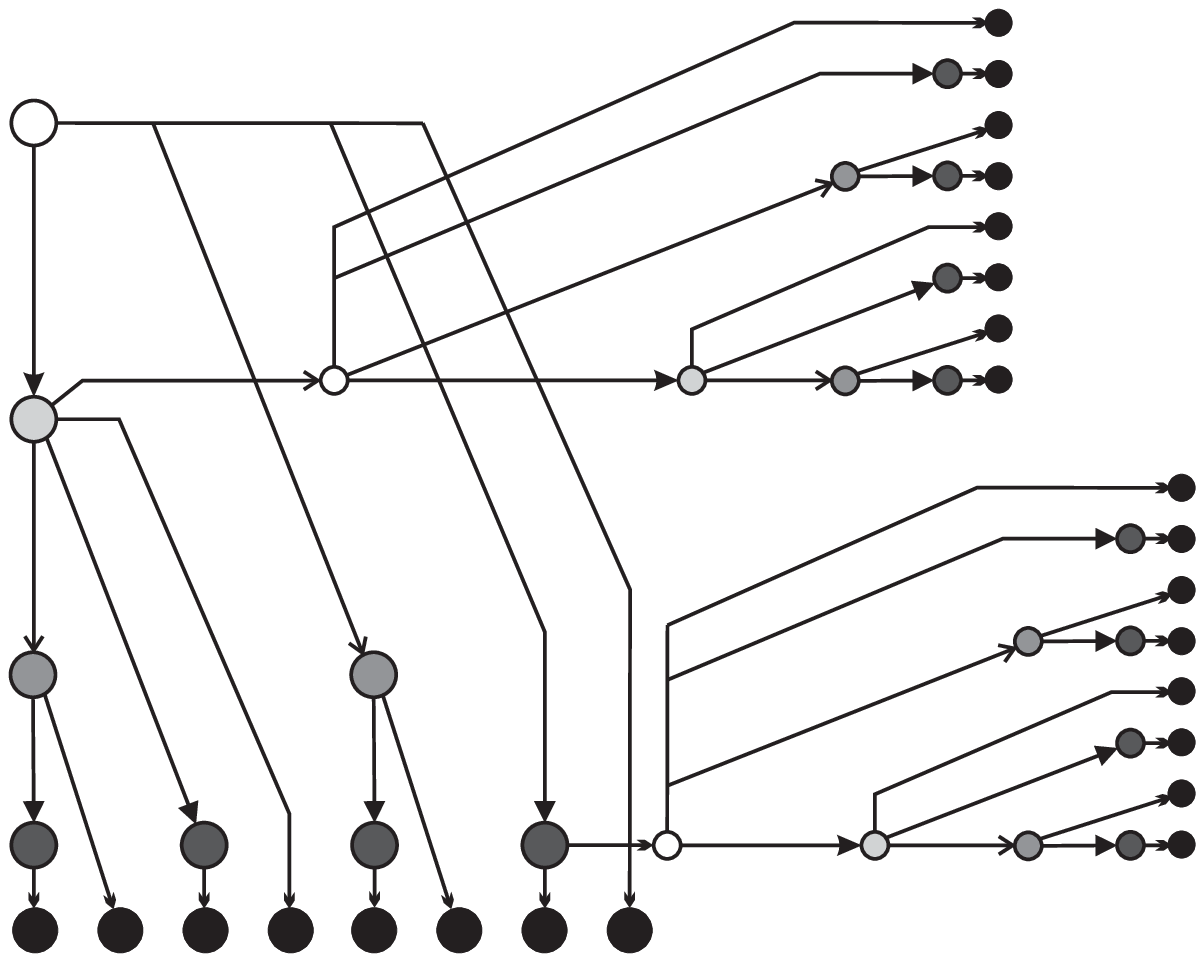}}
                           \parbox{.33\textwidth}{\center\includegraphics[width=.3\textwidth]{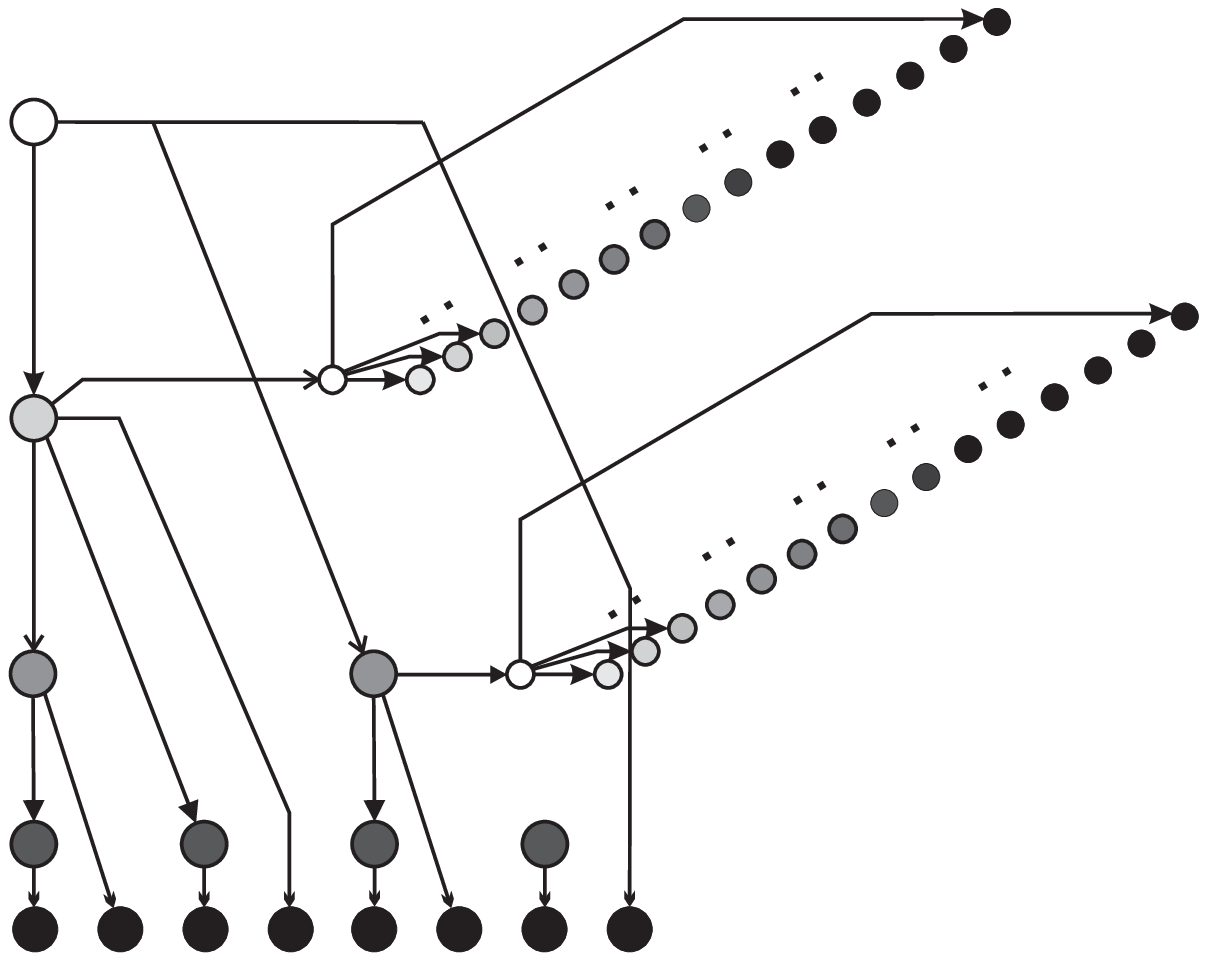}}}
       \parbox{\textwidth}{\parbox{.33\textwidth}{\center\scriptsize(a) SNAP plus mesh}
                           \parbox{.33\textwidth}{\center\scriptsize(b) SNAP in SNAP}
                           \parbox{.33\textwidth}{\center\footnotesize(c) OPST in SNAP}}
       \vspace{-2ex}
    \caption{Various ways of applying SNAP in P2P media streaming.}\vspace{-2ex}\label{fig:hybrid}
   \end{center}
  \end{figure*}
  Although SNAP delivers minimum streaming delay and scales well with an increasing number of peers, its high overhead in coping with peer dynamics along with its limited compatibility with bandwidth heterogeneity makes it impossible to build a P2P streaming system based solely on SNAP. This has been explained in Sec.~\ref{sec:peerdy} and \ref{sec:hetero}, where a hybrid system design has been suggested.

  In this section, we propose several extended system designs for P2P media streaming, which make use of SNAP as a component while complementing it with other mechanisms. Our system designs embrace a hybrid and hierarchical structure. In particular, we gather stable peers into SNAP to form a streaming backbone, and other peers simply attach themselves to the backbone in various ways (as we will explain below). A couple of excellent techniques have been proposed recently to identify stable peers \cite{WangLX-INFOCOM08,LiuWLZ-INFOCOM09}; we hence directly assume the availability of stable peers and refer the interested readers to the related literature for more details.

  We assume that a non-empty subset of the peers in SNAP backbone have surplus bandwidth, i.e., $B_i > B_\mathrm{base}$ for $i \in \widetilde{\mathcal{N}}_\mathrm{SNAP} \subseteq \mathcal{N}_\mathrm{SNAP}$, where $\mathcal{N}_\mathrm{SNAP} \subseteq \mathcal{N}$ is the set of all peers in SNAP. This assumption makes sense because, as $B_\mathrm{base}$ is set to the streaming rate, $B_i = B_\mathrm{base},~\forall i \in \mathcal{N}_\mathrm{SNAP}$ would suggest the impossibility to extend the system anymore. We illustrate several system designs in Fig.~\ref{fig:hybrid} and explain each of them separately. To simplify the illustration, we always use one tree to represent a multi-tree overlay in the figures.

\subsection{Hybrid Push-Pull Streaming}
  This a very nature design to coping with peers dynamics and bandwidth heterogeneity. As shown in Fig.~\ref{fig:hybrid}(a), those non-stable or low-bandwidth peers form a mesh to help each other in disseminating media chunks, and they also pull chunks from the backbone (the double-ended arrows indicate the bi-direction communications involved in the pull procedures between non-backbone and backbone peers). Some of the peers might have just joined the streaming system, hence their stability has not been tested yet. If they are later identified as stable and they do provide sufficient bandwidth, they will be added into the SNAP backbone.

  Chunk dissemination in SNAP is put in a higher priority. Therefore, peers in SNAP backbone will first use their bandwidth to meet the need for pushing chunks in the backbone. These peers respond to a pull request from non-backbone peers only if they surely have surplus bandwidth. We refrain from discussing the chunk scheduling strategies for the mesh-based pull mechanism used among the non-backbone peers, as these strategies have been intensively investigated; interested readers are referred to the literature provided in Sec.~\ref{sec:rw}. It is worth noting that this hybrid design may actually coexist with other designs below (which only involve different push schemes in a hierarchical manner), but we will not repeat the description of this hybrid design when discussing other designs.

\subsection{Hierarchical Push Streaming with SNAP in SNAP}
  There are cases where certain peers are stable but they cannot provide sufficient bandwidth because, for example, they are behind NATs or they are residential users having low bandwidth ADSL. For these peers, dragging them into the backbone with those bandwidth abundant and stable peers may only degrade the performance of the whole system; this consequence actually applies not only to SNAP but also to pure mesh-based pull systems.

  In our hierarchical push streaming design, we advocate the use of layered coding (LC) \cite{WienCGHA-TCSVT07} or multiple description coding (MDC) \cite{Goyal-SPMag01} to deliver media with gracefully degraded quality among these peers. More precisely, these peers form sub-SNAPs with a reduced baseline rate $B_\mathrm{base}$ that can be afforded by them. As shown in Fig.~\ref{fig:hybrid}(b), the sub-SNAPs are attached to peers in the backbone SNAP, and these backbone peers actually serve as the ``servers" for the sub-SNAPs. These ``servers" re-encode the chunks they received and push either the base layer of LC or one description of MDC to a sub-SNAP. While the peers in sub-SNAPs may definitely pull chunks from the backbone to improve the media quality, participating in sub-SNAPs at least allow them to enjoy smooth playback (though with lower quality) in a timely fashion.

\subsection{Hierarchical Push Streaming with OPST in SNAP}
  \begin{figure*}[ht]
   \begin{center}
       \parbox{\textwidth}{\parbox{.33\textwidth}{\center\includegraphics[width=.32\textwidth]{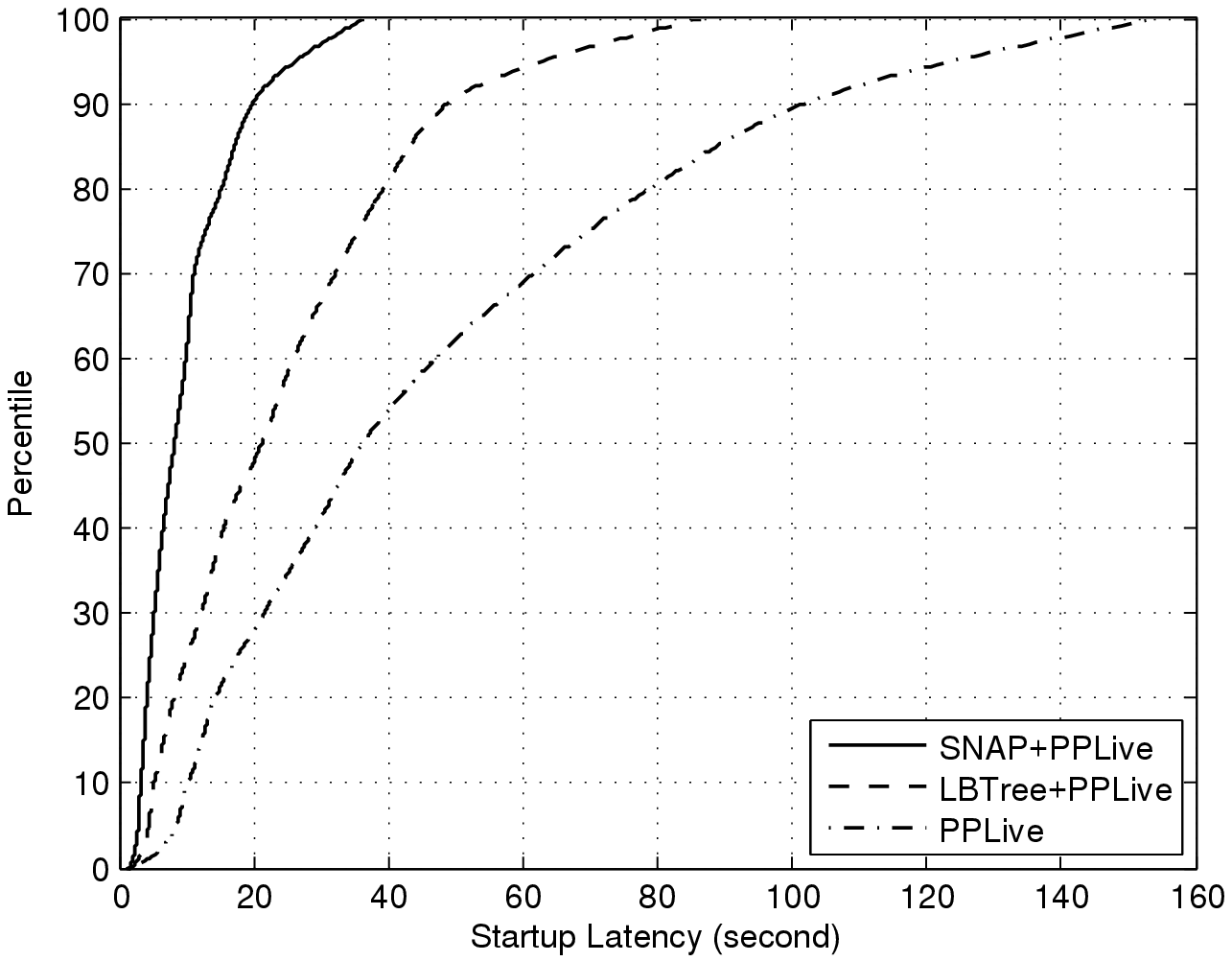}}
                           \parbox{.33\textwidth}{\center\includegraphics[width=.32\textwidth]{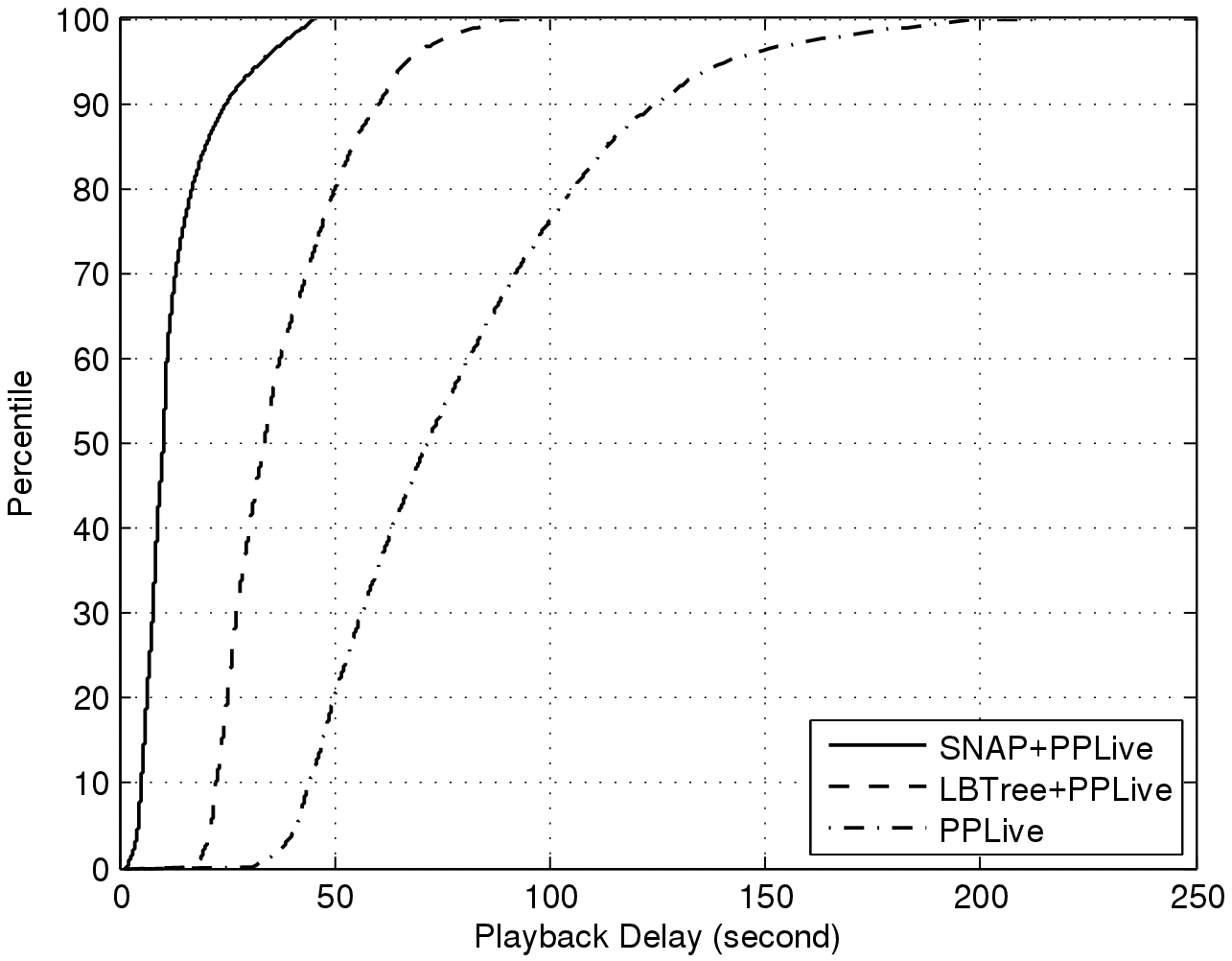}}
                           \parbox{.33\textwidth}{\center\includegraphics[width=.32\textwidth]{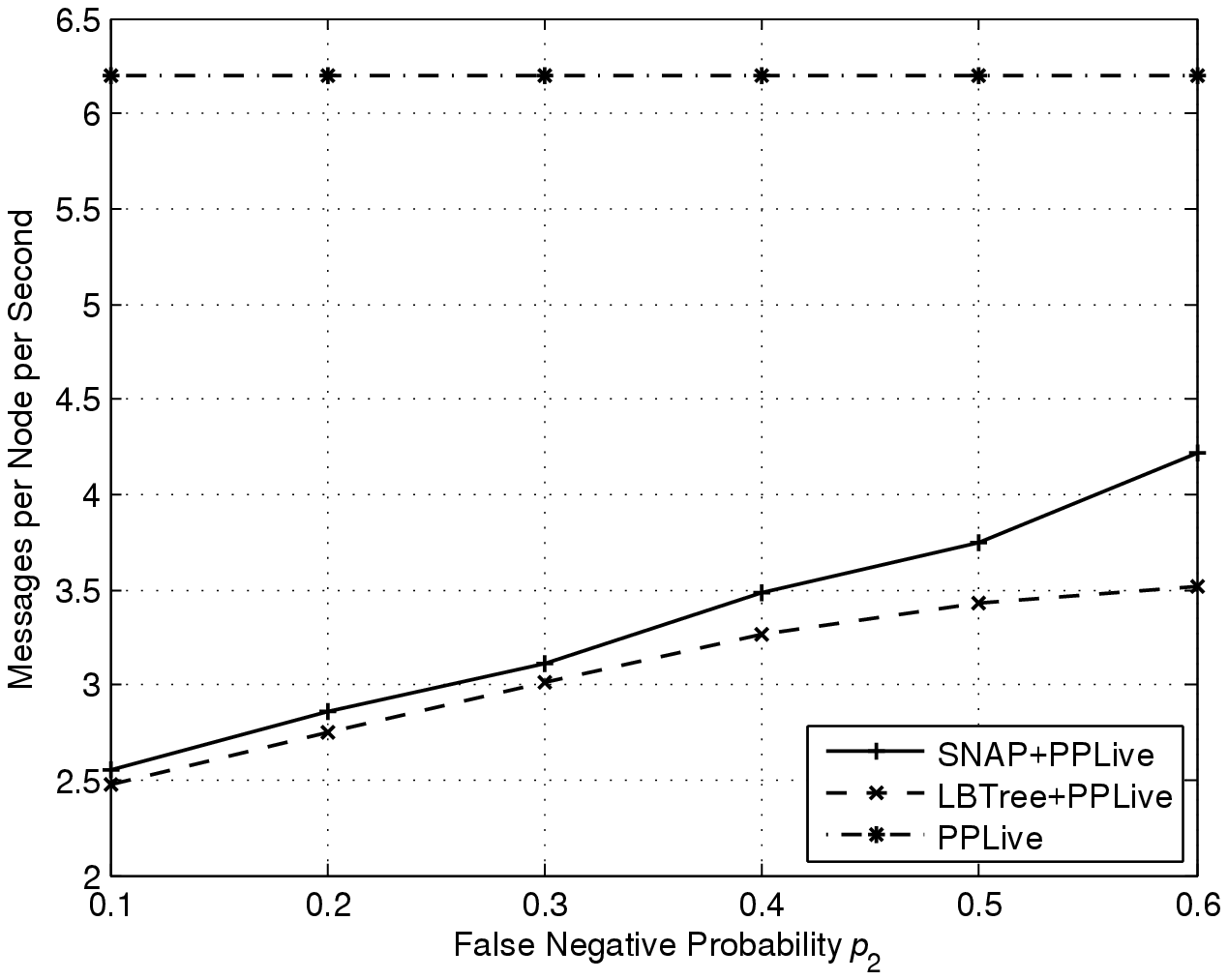}}}
       \vspace{-2mm}
       \parbox{\textwidth}{\parbox{.33\textwidth}{\center\scriptsize(a) CDF of startup  latency.}
                           \parbox{.33\textwidth}{\center\scriptsize(b) CDF of playback delay.}
                           \parbox{.33\textwidth}{\center\footnotesize(c) Control overhead as a function of $p_2$.}}
       \vspace{-2ex}
    \caption{Comparing SNAP+PPLive with LBTree+PPLive and PPLive. We only plot the mean values in (c), as the 95\% confidence interval is very narrow for every point in the figure due to the large amount of data being collected.}\label{fig:comp}
   \end{center}
   \vspace{-2ex}
  \end{figure*}

  As we have shown in Sec.~\ref{sec:best}, the superiority of SBT over OPST stems from the chunked nature of the media to be streamed; OPST could actually perform better if the media is packetized. However, the main obstacle of using OPST to stream packetized media (which we have briefly explained in Sec.~\ref{sec:intro}) is its high maintenance complexity and its incompatibility with a mesh-based pull mechanism. Nevertheless, OPST may still be applied in certain circumstances. For example, if the dissemination has reached the boundary of a LAN (wired or wireless) or a set of peers behind the same NAT, OPST actually becomes the best solution, as the maintenance becomes easy in such stub networks and there is no need to be combined with a mesh-based pull mechanism anymore.

  We illustrate such a design in Fig.~\ref{fig:hybrid}(c). Similar to SNAP in SNAP, the OPST in SNAP design let the localized peers form sub-OPSTs and attach themselves to some backbone peers that serve as ``servers" for these sub-OPSTs. These ``servers", after receiving media chunks, packetize them and stream the packets to the sub-OPSTs. Note that we use OPST simply as a representation for many other tree-based streaming techniques \cite{CastroDKNRS-SOSP03,VenkataramanYF-ICNP06,LiuSJRC-SIGMETRICS08} that equally apply here.

\section{Performance Evaluation} \label{sec:perfevl}
  To evaluation the performance of our proposed algorithms, we have implemented our SNAP in \textit{ns}-2. We have also implemented two other systems, PPLive-like system \cite{HeiLLLR-IPTV06} and LBTree+PPLive \cite{WangLX-INFOCOM08}, to be compared with ours. While PPLive is a typical mesh-based pull system, LBTree+PPLive is a hybrid system that combines a tree-based backbone with PPLive. We use the same parameter settings as those in \cite{WangLX-INFOCOM08}. For our system, we take SNAP+PPLive, i.e., we organize stable peers into SNAP and use PPLive to accommodate the remaining peers. We apply three metrics, \textit{startup latency}, \textit{playback delay}, and \textit{control overhead}, for the comparison. We refer to \cite{WangLX-INFOCOM08} for the definitions of these metrics. Note that, due to the performance limitation of \textit{ns}-2, our simulations are only performed for systems with hundreds of nodes. This is a weakness that will be overcome in our future implementations and experiments.
  
  The simulation configurations are as follows. The underlying topology is generated by GT-ITM \cite{itm}, where the intra-transit bandwidths are 2Gbps and the transit-stub/intra-stub bandwidths are set to 2Mbps, 5Mbps, and 10Mbps with equal probability. The link delay varies from 50ms to 500ms. Only nodes in stub-domains may participate P2P streaming. The session length is 3600 seconds and each chunk is 300Kb (1-second video), and we run 30 sessions for each system. We let 100 nodes joint the session at the beginning, the rest arrive following a Poisson process at rate 1 peer/second, and the maximum number of peers is 500. To focus more on the streaming performance, we did not implement the algorithm to identify stable peers. We use two probability $p_1=0.3$ and $p_2=0.1$ to characterize the behavior of stable nodes: whereas $p_1$ represents the fact that only 30\% of the arriving peers are identified as stable, $p_2$ captures the false negative of an identification algorithm. We use Pareto distribution (mean 100 seconds, $\alpha=1$) to model the durations of non-stable peers (including stable peers that are falsely identified), whereas a (real) stable peers stays until the end.

  In Fig.~\ref{fig:comp} (a) and (b), we plot the empirical \textit{cumulative distribution function} (CDF) of both startup latency and playback delay of all the three systems: SNAP+PPLive, LBTree+PPLive, and PPLive. It shows that, though LBTree+PPLive has already achieves a great improvement over PPLive (already shown by \cite{WangLX-INFOCOM08}), SNAP+PPLive obtains further improvement over LBTree+PPLive. We attribute this further improvement to the optimality of the SBT trees used by SNAP. As we have explained in Sec.~\ref{sec:best}, SBT performs better than the OPST given $t \approx d$, and the improvement becomes increasingly significant for larger $N$. As LBTrees are constructed by certain heuristics, we are not expecting too much difference between its performance and that of OPST. Note that this improvement does not only help the SNAP peers, the non-SNAP peers also have their delay performance improved accordingly due to the additivity of the delay. One may complain that the actually delay obtained through simulations are worse than the optimistic case where all peers are in the SNAP backbone (which should lead to a maximum delay of only a few seconds). However, there are two exceptions (to the optimistic case) in the simulation. Firstly, only 30\% of the peers are really stable peers, so the remaining peers have to be accommodated by the PPLive overlay, whose delay is significantly larger than that of SNAP. Secondly, as certain SNAP peers are falsely identified, they will leave after a short period and these peer leavings will lead to tree repairing (as explained in Sec.\ref{sec:peerdy}), which can also increase the delay.

  We also evaluate the control overhead for the three systems. Before looking into the simulation results, we first briefly discuss what constitute the control overhead of the three systems. For PPLive, the overhead mostly comes from (i) regular gossiping to maintain the mesh overlay, (ii) regular chunk availability advertisements, and (iii) chunk requests. Both SNAP+PPLive and LBTree+PPLive have the above three components in their control overhead for non-backbone peers. However, those non-backbone peers that interface the two tiers can have their overhead suppressed, thanks to the stability of their upstream peers in the backbone. The alternative overhead brought by SNAP and LBTree is to maintain the backbone. This part of the overhead varies with the stability of the backbone. In our simulation, we vary the value of $p_2$ to emulate the stability changes of the backbone: the smaller the value, the more stable the backbone is. As shown in Fig.~\ref{fig:comp} (c), the control overhead of PPLive remains constant with different values of $p_2$, while that of SNAP+PPLive and LBTree+PPLive naturally increases with $p_2$. Also, LBTree+PPLive performs constantly better than SNAP+PPLive, but the discrepancy in overhead is negligible for small values of $p_2$. This is reasonable as SNAP does entail more efforts to maintain its multi-SBT overlay, compared with the loosely coupled multi-tree structure of LBTree. Note that the large discrepancy takes place only when more than 50\% of the backbone peers are incorrectly identified, which are very pessimistic cases. Therefore, we deem our SNAP an excellent mechanism to trade overhead for better delay performance.

\section{Related Work} \label{sec:rw}
  The algorithms for P2P media streaming can be roughly categorized into two groups: namely \textit{push} and \textit{pull}. Whereas a media transmission is initiated by the sender in a push method, it is the receiver that requests specific media content first in a pull method. We discuss these two methods as well as their combinations in the following. Due to the space limitation, we only focus on algorithm aspects but omit the literature survey on system modeling and analysis.

  \textit{Structured push} is the most direct method for P2P media streaming; it involves either a single source-rooted tree \cite{ChuRZ-SIGMETRICS00,BanerjeeBK-SIGCOMM02} or multiple trees \cite{CastroDKNRS-SOSP03,VenkataramanYF-ICNP06,LiuSJRC-SIGMETRICS08}. This method inherits from the IP multicast \cite{DeeringC-TCS90} but brings the logical structure up to the overlay; the packets of a given media content are streamed along the paths predefined by the overlay structure. If the peers are all stable, a structured push delivers very good (even optimal if properly engineered \cite{LiuSJRC-SIGMETRICS08}) performance. However, its robustness against peer dynamics is limited, and any attempt to improve it may lead to very high complexity \cite{MaghareiRG-INFOCOM07}.

  A complementary method is the \textit{mesh-based pull} \cite{KosticRAV-SOSP03,ZhangLLY-INFOCOM05,MaghareiR-INFOCOM07,HeiLR-IEEECOMMAG08,LiXQKLLZ-INFOCOM08,FengLL-INFOCOM09}. It replaces the structured overlay with a loosely coupled mesh, in which the neighborhood relation is maintained by, for example, gossiping \cite{KosticRAV-SOSP03,ZhangLLY-INFOCOM05}. The benefit is the great improvement in coping with peer dynamics. However, the removal of the structured overlay ``blinds" all the peers in a sense that a peer has no idea which part of a media content needs pushing out and where to push it. Consequently, peers need to exchange buffer maps periodically and they request any missing pieces of a media content from others based on the received buffer maps. In order to make the protocol scalable, the media content has to be represented by a unit much larger than packet, which is usually referred to as chunk and has a size of several hundreds kilobytes. As a new chunk can be requested only upon the exchange of new buffer maps, the pull approach suffers from the inherent tradeoff between protocol efficiency and chunk delay.

  \textit{Randomized push} \cite{MassoulieTGR-INFOCOM07,BonaldMMPT-SIGMETRICS08} attempts to improve the robustness of the structured push by randomizing the peers' view on their neighbors and to suppress the overhead of the pull method by avoiding the receiver requests. Unfortunately, as the buffer map exchange is still necessary, the fundamental tradeoff between efficiency and delay is inevitable.

  The fact that the push and pull methods have different but complementary pros and cons has motivated several recent proposals on combining them into a hybrid system \cite{WangXL-ICDCS07,ZhangZSY-IEEEJSAC07,WangLX-INFOCOM08}. Whereas these existing proposals construct the backbone trees using heuristics, we are the first to come up with a multi-tree backbone that achieves the optimum delay performance. As shown by our simulation results in Sec.\ref{sec:perfevl}, it is definitely beneficial to make use of the optimal overlay construction.

\section{Conclusion} \label{sec:con}
  In this paper, we have investigated the issue of building a hybrid and hierarchical P2P streaming system. We have focused on the design of the streaming backbone that consists of stable peers. Based on the theoretical results of \cite{Liu-MM07} (where the existence of a minimum delay scheduling policy is shown), our SNowbAll multi-tree Pushing (SNAP) applies a distributed chunk scheduling policy guided by a multi-tree overlay we propose, and it guarantees minimum delay of chunk streaming. We use SNAP as a backbone to organize the stable peers, and we also present various way of combining SNAP with other overlay structures at the second tier to accommodate other peers. Using simulations with \textit{ns}-2, we have demonstrated the effectiveness and efficiency of SNAP.

  Inspired by the exciting results from our simulations, we are planning to deploy such a streaming system in NTU campus. In NTU, we have a lot of online teaching programs to allow students learning at home. Currently, these online contents are distributed through a centralized server. As the available time period for each learning program is pretty short, the access to it is very much synchronized. This makes P2P streaming a very attractive way to release the burden of the server, as well as to improve the playback quality. Also, the synchrony implies that, should a P2P streaming system be used, the peers would almost be stable, hence SNAP is a perfect solution to organize the peers. This deployment will help us to further enhance the design of SNAP.

\bibliography{SNAP}
\bibliographystyle{IEEEtran}

\appendices
\section{The Average Delay of The Snowball Tree} \label{sec:avg}
   We first assume $N=2^K$ for some integer $K$. According to the assumption we made in Sec.~\ref{sec:best}, the peer in the 0-th level has a delay $D_0 = 0$ and the peer in the 1-st level has a delay $D_1 = \bar{d}+t$. For the $k$-level where $k \geq 2$, there are more than one peers and they may experience different delays according to how much pipelining is involved in the transmission. We denote by $D_{k,\mathrm{min}}$ and $D_{k,\mathrm{max}}$ the minimum and maximum delay among all the delays experienced by peers in the $k$-th level. Obviously, we have $D_{k,\mathrm{min}} = \bar{d} + kt$ and $D_{k,\mathrm{max}} = k\left(\bar{d} + t\right)$. Now we show that the average delay in the $k$-th level is
   $\bar{D}_k = \frac{D_{k,\mathrm{min}} + D_{k,\mathrm{max}}}{2} = \frac{k+1}{2}\bar{d} + kt$.

   The proof is by induction. First, it is easy to verify that the equation holds for $k=3$. For $k>3$, we have
   \begin{eqnarray}
   \!\!\bar{D}_k
   \!\!\!\!&=&\!\!\!\! \frac{\sum_{i=0}^{k-1} 2^{i-1}\left[\bar{D}_i + \left(\bar{d}+(k-i)t\right)\right]}{2^{k-1}} \label{eq:avgdelay}\\
   \!\!\!\!&=&\!\!\!\! \frac{(d+kt)+\displaystyle{\sum_{i=1}^{k-1}} 2^{i-1}    \left[\left(\frac{i+1}{2}\bar{d}+it\right)+\left(\bar{d}+(k-i)t\right)\right]}{2^{k-1}} \nonumber \\
   \!\!\!\!&=&\!\!\!\! \frac{k+1}{2}\bar{d} + kt \nonumber
   \end{eqnarray}
   The equality (\ref{eq:avgdelay}) follows from the fact that, if a peer in the $k$-th level receives a chunk from some peer in the $i$-th level ($i<k$), it suffers an extra delay of $\bar{d}+(k-i)t$ due to the pipelining effect. As a result, the overall average for the whole system is
   \begin{eqnarray}
   \bar{D}^{\mathrm{SBT}}
   &=& \frac{\textstyle{\sum_{i=0}^{K}} 2^{i-1} \bar{D}_i}{2^K}~~=~~\frac{\sum_{i=1}^{K} 2^{i-1} \left(\frac{i+1}{2}\bar{d}+it\right)}{2^K} \nonumber \\
   &=& \frac{K}{2}\bar{d} + (K-1)t + \frac{t}{2^K} \nonumber
   \end{eqnarray}
   For $\lceil\log_2 N\rceil - 1 < K < \lceil\log_2 N\rceil$, we could always schedule the peers in the last level to those paths with smaller delays. Therefore, the average delay for $K=\lceil\log_2 N\rceil$ serves as an upper bound of those for $\lceil\log_2 N\rceil - 1 < K < \lceil\log_2 N\rceil$.

\section{The Characterization of Feasible Tree Edge Scheduling: Proposition~\ref{prop:feasible}} \label{sec:tsch}
  To prove the two iff conditions, we first need to understand that two edges can upload concurrently iff they do not share origins and both of their origins have chunks to upload. The sufficiency is pretty trivial to see, while the necessity is the direct consequence of applying sequential uploading at every peer (whose optimality is suggested by \cite{Liu-MM07}): as parallel uploading is not allowed, a peer cannot perform more than one uploading at a time.

  Now, let us first look at the common tree case. The sufficiency actually stems from the construction of SBTs and the definition of an edge level set: the origins of an edge level set are all different and the edges in such a set are meant to upload the same chunk available at the edge origins. The necessity follows directly from the causal ordering of the level sets: the edges in $L^e_{k,i}$ cannot be scheduled before those in the level sets whose indices are small than $k$, as otherwise the origins of them may either have not received the chunk to be disseminated or be uploading the chunk through edges in the level sets whose indices are small than $k$.

  As for the distinct tree case, the requirement of non-common origins is easy to understand, but to prove another condition about indices difference is slightly more tricky. This condition follows from both the causality within a SBT and time sequence among consecutive SBTs. In particular, $i-j=a~(\!\!\!\!\mod P)$ implies that the chunk pushed to $T_i$ is $a$ time slots later than that pushed to $T_j$. Now, if the uploading in $T_i$ has progressed till the $k_1$-th level, the uploading in $T_j$ cannot go beyond $k_2$-th level for $k_2 = k_1 + a$, as it would otherwise violated the causality in $T_j$ (a similar reason accounting for the common tree case). Conversely, if we have $k_2 \leq k_1 + a$, then we are sure that the origin of $l_2$ has a chunk available to send. \hfill Q.E.D.

\section{Sufficient Condition for Minimum-Delay Multi-Tree Overlay: Proposition~\ref{prop:sufficient}} \label{sec:mdmtree}
  Suppose that we do schedule all the possible edges stated in Proposition~\ref{prop:sufficient} at a given time slot. As the edge in $L^e_{1,i}$ is uploading in $T_i$, the server has pushed a chunk to $T_i$ in the previous time slot. According to the time sequence among all the SBTs, the server pushed to $T_{i-K+1}$ a chunk $K$ time slots before. Since the number of edges in $L^e_{K,i-K+1}$ that are uploading, $N-2^{K-1}$, is sufficient to complete the chunk dissemination for $T_{i-K+1}$, it is straightforward to see that the minimum delay of $K$ time slots is achieved for the chunk disseminated by $T_{i-K+1}$. However, as $T_i$ (hence $T_{i-K+1}$) is arbitrarily chosen, the result actually applies to every SBT in the multi-tree structure. \hfill Q.E.D.

\vfill\eject
\section{Correctness of The Centralized Multi-SBT Construction Algorithm} \label{sec:algomtree}
  One can easily verify that the two algorithms presented in Fig.~\ref{fig:algomtree} and Fig.~\ref{fig:algomtreeplus} are tailored to the conditions stated in \textit{Proposition~\ref{prop:feasible}} and \textit{Proposition~\ref{prop:sufficient}}. In particular, the periodic allocations of the peers in the multi-SBT structure make sure that all the peers in an ISet are distinct and the level sets scheduled for neighboring SBTs differs exactly by one, which satisfies the last condition stated in  \textit{Proposition~\ref{prop:feasible}} and the sufficient condition required by \textit{Proposition~\ref{prop:sufficient}}. Therefore, as far as we can show the algorithm terminates correctly, the resulting multi-SBT structure will meet our need: it will guarantee the minimum delay for chunk streaming.

  Actually, proving the termination of the extension algorithm in Fig.~\ref{fig:algomtreeplus} is trivial: as it only manages to allocate the extra $N - 2^{\log_2\lfloor N \rfloor}$ peers into the same number of positions, the algorithm is bounded to terminate. Therefore, the key is to prove the termination for the basic algorithm shown in Fig.~\ref{fig:algomtree}. More precisely, we need to show that $2^K$ peers are enough to fill up the ISet $\{s_0, s_1, \cdots, s_K\}$. Since we have $|s_k| = (K-k)2^{(k-1)^+}$, the total number of peers needed to build an ISet is given by $\sum_{k=1}^K (K-k)2^{(k-1)^+} = 2^K - 1$. As there are in total $2^K$ peers, the termination of the algorithm is guaranteed.  \hfill Q.E.D.

\section{Bounding the Neighbor Table Size: Proposition~\ref{prop:nbsizebd}} \label{sec:bndntb}
  Consider a peer at the $k_1$-th level and its children at the $k_2$-th level. The worst case is that $P_{k_1}$ and $P_{k_2}$ are \textit{coprime}, in which case the neighbor table contains $P_{k_2}$ peers at the $k_2$-th level. Therefore, a peer may have the largest neighbor table only if it belongs to the 0-th level and $P_0$ is coprime with $P_1,\cdots,P_{K-1}$, and the maximum size is $1+\sum_{k=1}^{K-1} k = 1+\frac{1}{2}K(K-1)$, where the extra one comes from the $K$-th level (whose period, $P$, is never coprime with $P_0$). \hfill Q.E.D.

\end{document}